\documentstyle [aps,prc,preprint]{revtex}
\begin {document}
\draft
\title
{Chirally Constraining the $\pi \pi$ Interaction in Nuclear Matter}
\author
{R. Rapp$^{1}$, J.W. Durso$^{2,1}$, and J. Wambach$^{1,3}$}
\address
{1) Institut f\"ur Kernphysik, Forschungszentrum J\"ulich,
    D-52425 J\"ulich, Germany \\
 2) Physics Department, Mount Holyoke College, South Hadley,
 MA 01075, U.S.A.\\
 3) Department of Physics, University of Illinois, Urbana,
 IL 61801, U.S.A.}

\maketitle

\begin{abstract}
A general prescription for the construction of $\pi \pi$ interaction
potentials which preserve scattering length constraints from chiral
symmetry when iterated in scattering equations is derived.  The
prescription involves only minor modifications of
typical meson-exchange
models, so that coupling constants and cut-off masses in the models
are not greatly affected.
Calculations of  $s$-wave $\pi \pi$ scattering amplitudes in
nuclear matter for two models are compared with those for similar
models which violate the chiral constraint.  While the prescription
tends to suppress the accumulation of the near sub-threshold strength
of the $\pi \pi$ interaction, an earlier conjecture that amplitudes
which satisfy chiral constraints will not exhibit an
instability towards
$\pi \pi~s-$wave pair condensation appears to be incorrect.  At the
same time, however, conventional $\pi \pi$ interaction models which
fit scattering data well can readily be adjusted
to avoid the instability
in nuclear matter without recourse to exotic mechanisms.
\end{abstract}

\pacs{ }

\section* {1. Introduction}

The role of correlated two-pion exchange in the interaction of two
nucleons is well-established \cite{Hol},
with the bulk of the intermediate\-
range attraction due to the scalar-isoscalar
($\sigma$-meson) channel.  This
same mechanism is presumably responsible for
most of the attraction between
nucleons in nuclei, despite the possibly
strong influence of the nuclear
medium on the $\pi \pi$ correlations \cite{Eri}, especially in the
$\sigma$-channel \cite{Sch}.  There the coupling of
pions to nucleons and
$\Delta\/$'s enhances the attraction of the pions at
energies near the 2$\pi$
threshold, which can lead to quasi-bound states, and even to $\pi \pi$
$s$-wave pair condensation at relatively low densities \cite{Cha}.

In a recent study Aouissat {\em et al.\/} \cite{Aou}, hereafter referred
to as I, demonstrated the unattractive possibility of $\pi \pi$ pair
condensation at nuclear densities of 1.2-1.3$\rho _{0}$ ($\rho _{0}~=~
0.16~{\rm fm}^{-1}$, normal nuclear matter density)
for two phenomenological
models \cite{Lee,Loh} which fit $\pi \pi$ s-wave phases over a broad
energy range.  It was also found in I that the tendency towards
$\pi \pi$ pair condensation appeared to be dramatically suppressed if
the $\pi \pi$ amplitude were forced to satisfy
a constraint required by chiral symmetry, which is that the amplitude
must satisfy the so-called soft-pion theorems \cite{Wei}.
Implicit in this constraint is that the $\pi \pi$
s-wave scattering lengths be proportional to the pion mass as the
pion mass approaches zero.

Their study immediately suggests two questions: (1) Does the imposition
of the chiral constraint on the amplitude provide a guarantee against
$\pi \pi$ pair condensation in the nuclear medium at low or moderate
density? (2) How can one construct interactions based on meson-exchange
models which are unitary and which satisfy the
soft-pion theorems without
recourse to reformulating the treatment of pions in nuclear matter
\cite{Aou}? We will address these questions in reverse order
and first derive a simple
prescription for how to construct pseudopotentials in which the
scattering length constraint is guaranteed.  Having done this, we will
examine  reasons why meson-exchange models cannot
follow the prescription and
then address the problem of sufficiency posed in the first question.

\section*{2. Minimal Chiral Constraints}

The difficulty in imposing the constraint
on the scattering lengths, hereafter called the minimal
chiral constraint, or MCC, is that the $\pi \pi$ amplitudes used in the
in-medium calculations are generated from an interaction kernel which
is unitarized by means of a scattering equation ({\em e.g.\/}, the
Blankenbecler-Sugar equation \cite{Bla}).  In that case, any symmetries
of the interaction kernel are usually destroyed by iteration in the
scattering equation.  This is a well known result.  In meson-exchange
models, for example, the interaction kernel is the Born amplitude of
an effective field-theoretic lagrangian, which is crossing-symmetric.
Solving the scattering equation effectively
sums a subset (ladder/bubble graphs)
of the full set of Feynman diagrams, thereby destroying the crossing
symmetry of the kernel.  In the same way an interaction kernel which is
chirally symmetric in the limit of zero pion
mass results in an amplitude
in which this property is lost through iteration in the scattering
equation.

Were the objective simply to calculate scattering amplitudes for pions
at zero density, the problem of preserving (broken)
chiral (and crossing)
symmetry is solved, in principle, by chiral perturbation theory.
However calculations in chiral perturbation theory beyond one loop are
overwhelming, so that unitarity cannot easily be enforced, nor can one
readily adapt the method to the problem of pion interactions in the
nuclear medium.  For practical reasons we are led back to effective
models used with scattering equations.

\subsection*{2.1 Enforcing the MCC}

The problem of enforcing the MCC was solved in I by using a
once-subtracted form of the 2-pion propagator in the
Blankenbecler-Sugar (BbS)
equation \cite{Bla}, taking the subtraction point at $s=0$ ($s$ is the
square of the center-of-mass energy) and setting the subtraction
constant to zero.  For free particle scattering, this resulted in the
replacement of the BbS propagator
\begin{equation}
G_{\pi \pi }(s,k) = \frac{1}{\omega _{k} (s-4\omega _{k}^{2} + i\eta )}
\end{equation}
by the form
\begin{equation}
\tilde G_{\pi \pi }(s,k) = \frac{s}{\omega _{k} (s-4\omega _{k}^{2}
+ i\eta )  4\omega _{k}^{2}}~,
\end{equation}
where $k$ is the modulus of the c.m. 3-momentum of the pions and
$\omega _{k}= \sqrt{k^{2}+m_{\pi }^{2}}$.  The appearance of $s$ in the
numerator of $\tilde G_{\pi \pi }(s,k)$ as an {\em external\/} variable
in the integral equation ensures that the scattering amplitude generated
by solving the BbS equation (quantum numbers suppressed)
\begin{equation}
\label {BbS}
M(s;q,'q)=V(s;q',q)+\int_{0}^{\infty} k^{2}dk V(s;q',k)
\tilde G_{\pi \pi}(s,k) M(s;k,q)
\end{equation}
will be ${\cal O}(s)$ if $V(s;q',q)$ is ${\cal O}(s)$ as $q,q' \to 0$,
assuming that the integration over intermediate momenta gives a
factor of ${\cal O}(1)$ or
higher.  Since the scattering length in the $i$-th channel (where
$i$ stands for ($I=0,J=0$) or ($I=2,J=0$)) is given by

\begin{equation}
a^{i}=\lim_{q \to 0} \frac {M^{i}(s;q,q)}{32\pi \sqrt{s}},
\end{equation}
with $q=\sqrt{\frac {s}{4}-m_{\pi}^2}$, if $M^{i} \propto s$, then
$a^{i} \propto \sqrt{s}  \to 2m_{\pi}\/$ and the MCC is fulfilled.
The difficulty with this
solution is that the suppression of the rescattering terms at small
momenta forces one to use large coupling constants and unrealistically
high values of cutoffs, and to rely on heavy meson-exchange ({\em e.g.},
the $f_2$(1270)), in meson exchange models such as
the J\"ulich model \cite{Loh} in order to achieve a good fit
to $\pi \pi$ phase shifts, so that one loses contact with the
underlying low-energy physics of the $\pi \pi\/$ interaction.
Here we wish to construct
interaction kernels which will yield amplitudes which
satisfy the MCC without
having to resort to the subtraction scheme of I, and thereby avoid its
unattractive features.  In that sense, this is a continuation of the
work begun in I.  For that reason we will adopt the same model for
renormalizing the single-pion propagator in nuclear matter as in that
work. Since the model is described in detail in I, we will not repeat it
here.

The solution to building the MCC into the interaction kernel is quite
straightforward.  We noted earlier that the reason the subtracted form
of the 2-pion propagator preserved the scattering length constraint
contained in the interaction kernel is that $s\/,$ the square of the
center-of-mass energy, appears as an {\em external} variable in the
rescattering term in the integral equation, eq.~(\ref{BbS}).  Thus, if
$V(s;q,q)$ is ${\cal O}(s)$ as $q \to 0\/$, the same will hold for
$M(s;q,q)\/$.  If the kernel is such that it always contains
{\em external\/} kinematic factors which are each ${\cal O}(s)$ or
higher at each stage in the {\em iterative\/} solution of the scattering
equation, then the MCC will be satisfied.

One simple way to accomplish this is through the use of separable
potentials.  If $V(s;q',q)$ is a sum of terms of the form
$u_i(s,q')u_i(s,q)$ such that $u_i(s,q)\/$ is ${\cal O}(\sqrt{s})\/$
(or higher) as $q \to 0\/,$ the amplitude will be ${\cal O}(s)\/$
and the MCC will be fulfilled.  This property was exploited in I
with the linear $\sigma$-model. They started  from the $I=0$
Born amplitude in the linear $\sigma$-model,
\begin {equation}
M_{B}^{I=0} = \frac {m_{\sigma}^{2}-m_{\pi}^2}{f_{\pi}^{2}}
              \left( 3 \frac {s-m_{\pi}^{2}}{s-m_{\sigma}^{2}} +
                 \frac {t-m_{\pi}^{2}}{t-m_{\sigma}^{2}} +
                 \frac {u-m_{\pi}^{2}}{u-m_{\sigma}^{2}} \right),
\end {equation}
where $s\/$, $t\/$, and $u$ are the usual Mandelstam variables for
{\em physical\/} processes given by
\begin {eqnarray}
s & = & (q_1+q_2)^{2}=(q_3+q_4)^{2}  \nonumber \\
t & = & (q_1-q_3)^{2}=(q_4-q_2)^{2}   \\
u & = & (q_1-q_4)^{2}=(q_3-q_2)^{2}  \nonumber
\end {eqnarray}
with $q_1,~q_2$ the incoming and $q_3,~q_4$ the outgoing pion 4-momenta.

In order to construct a separable potential, they took the off-shell
continuation of $s\/$, $t\/$, and $u$ to be
\begin{eqnarray}
\label{offshell1}
s & = & E^{2}  \nonumber \\
t & = & 2m_{\pi}^{2}-2\omega _{q} \omega _{q'} + {\bf q \cdot q'} \\
u & = & 2m_{\pi}^{2}-2\omega _{q} \omega _{q'}-{\bf q \cdot q'},
\nonumber
\end{eqnarray}
corresponding to placing the pions on the mass shell
($\omega _{q} \equiv \sqrt{{\bf q}^{2}+m_{\pi}^{2}}\/$)
in $t$ and $u\/$,
but taking $s$ as an external variable in the scattering equation.
Neglecting the $t$ and $u$ in the denominators as being small compared
with $m _{\sigma}^{2}\/$, the interaction kernel becomes

\begin{equation}
\label{sepsigma}
V^{00}(E;q',q)=\frac {m_{\sigma}^{2}-m_{\pi}^{2}}{f_{\pi}^{2}}
 \left( 3 \frac {E^{2}-m_{\pi}^{2}}{E^{2}-m_{\sigma}^{2}}
  + \frac {4 \omega _{q} \omega _{q'} - 2m_{\pi}^{2}}{m_{\sigma}^{2}}
  \right)  v(q')v(q).
\end{equation}
Here $q$ and $q'$ are the moduli of ${\bf q}$ and ${\bf q'}$ and
$v(q)$ is a dipole-type form factor
with $v(0)=1$ to insure convergence of the integral.  Clearly
the interaction defined in eq.~(\ref{sepsigma}) satisfies the criteria
for MCC set out above: Every term contains either an external factor
({\em e.g.\/}, $E^{2}$ or $m_{\pi}^{2}$) or a factor which is a product
of terms ($\omega _{q} \omega _{q'}$) which survive the integration
and are ${\cal O}(m_{\pi}^{2})$ as $q,q' \to 0$.

Notice, however, that separability of the potential
is not a crucial part of the recipe
for ensuring MCC.  If it were, then all meson-exchange models would be
irreparable.  The important point is that all the terms in the potential
have in their numerators products of factors which are of order $m_\pi$
in the chiral limit, one of which depends only on the initial momentum
and the other only on the final momentum.  Were the $t$- and
\protect {$u$-dependence} of the denominators retained in defining
$V^{00}(E;q',q)$, the MCC would still be satisfied.  This is evident
if one writes the scattering equation in iterated form (schematically):
\begin {equation}
M~=~V~+~VGV~+~VGVGV~+\ldots
\end {equation}
Every term in the series will contain one factor which depends on the
initial momentum, and one which depends on the final momentum.  As long
as the integrals implied in the equation are ${\cal O}(1)$ or higher
(the form factors used in the potential guarantee this), the MCC is
satisfied by $M$.  Thus, it is the choice of off-shell continuation
of the kinematic variables, and not separability, which is crucial to
preserving the scattering length constraint.  Observe also that the
choice of off-shell continuation is not unique.  We could have taken
$s~=~(q_{1}+q_{2})^{\mu}(q_{3}+q_{4})_{\mu}~=~4\omega _{q} \omega _{q'}$
and still fulfilled the prescription for satisfying the MCC.  The
choice $s=E^{2}$ is made on physical grounds: we want the $\sigma$-meson
to appear as a fixed pole in energy in the scattering channel.

A generalization of the prescription for construction of an
MCC-preserving potential is now clear:
\begin{itemize}
\begin{enumerate}

\item Start with an on-shell Born amplitude derived from a chirally
symmetric lagrangian (broken by the pion mass term), expressing it
in terms of the Mandelstam variables in crossing-symmetric form.

\item Make the off-shell continuation of $s$, $t$, and $u$ as indicated
in eq.~(\ref{offshell1}).

\item Include form factors, as necessary, to insure convergence of the
rescattering term.

\end{enumerate}
\end{itemize}
We will refer to potentials of this type as MCC quasi-separable.

Of course, the above prescription is sufficient, but not necessary.  It
is certainly possible that potentials can be constructed which are not
of this type but which, through a delicate cancellation of terms, do
fulfill the MCC.  Nevertheless, the prescription allows us to define,
for any underlying effective lagrangian which is (broken) chirally
symmetric, a potential which will yield an amplitude that has the
proper behavior of the scattering lengths.

\subsection*{2.2 Inconsistencies in the MCC Prescription}

Although the prescription above guarantees MCC-compliance of the
interaction kernel, it is neither unique, nor free from
inconsistencies which seem to be unavoidable.  We have
already alluded indirectly to such inconsistencies
in our discussion above
of the relation of the continuation of the Mandelstam variables via
eq.~(\ref{offshell2}) to the BbS equation.
These appear to arise whenever
one uses an effective lagrangian to generate a Born term to use in a
scattering equation; that is, the off-shell continuation needed to
satisfy the MCC is not derivable from the evaluation of interaction
vertices based on a lagrangian used consistently
with a scattering equation.

We illustrate this with a common example: the exchange of a
$\rho$-meson.  The interaction term in the lagrangian is
\begin{equation}
{\cal L}_{\rho\pi\pi}=-g_{\rho}(\vec \pi \times \partial ^{\mu} \vec
\pi) \cdot \vec \rho _{\mu}~.
\end{equation}
The $I=0$ and $I=2$ Born amplitudes contain $t$-
and $u$-channel exchange  terms which are given by
\begin{equation}
\label{rhob}
M^{I=0,2}_{\rho ,B} = (2,-1) 2g_{\rho}^{2}
\left[ \frac {(q_{1}+q_{3})^{\mu}
(q_{2}+q_{4})_{\mu}}{(q_{1}-q_{3})^{2}-m_{\rho} ^{2}} +
\frac {(q_{1}+q_{4})^{\mu}q_{2}+q_{3})_{\mu}}
{(q_{1}-q_{4})^{2}-m_{\rho} ^{2}} \right].
\end{equation}
We have omitted the contributions arising from gauge terms in the $\rho$
propagator, since they vanish for the fully on-shell, equal-mass Born
amplitude, and they are usually neglected in the meson-exchange
potentials as well.  In that case the amplitude, expressed in manifestly
crossing-symmetric form, is
\begin{equation}
\label{rhob2}
M^{I=0,2}_{\rho ,B} = (2,-1) 2 g_{\rho}^{2}
\left( \frac {s-u}{t-m_{\rho} ^{2}}
+ \frac {s-t}{u-m_{\rho} ^{2}} \right).
\end{equation}
Let us examine this potential as evaluated {\em consistently}
in the BbS approach, and then in time-ordered perturbation theory.

In the BbS approach, the interaction is instantaneous, so that only
3-momentum is transferred in the exchange of the meson.  Thus the
scattering particles remain on the energy shell, but not on the mass
shell.  In the c.m. frame $q_{1,2}=(E/2,(+,-){\bf q})$ and
$q_{3,4}=(E/2,(+,-){\bf q'})$.  In terms of these variables,
eq.~\ref{rhob} becomes
\begin{equation}
M^{I=0,2}_{\rho ,B} = (2,-1) 2 g_{\rho}^{2}
\left( \frac {E^{2}+q^{2}+q'^{2}+2{\bf q \cdot q'}}
{-({\bf q-q'})^{2}-m_{\rho} ^{2}}
+ \frac {E^{2}+q^{2}+q'^{2}-2{\bf q \cdot q'}}
{-({\bf q+q'})^{2}-m_{\rho} ^{2}} \right).
\end{equation}
This form can be obtained from the off-shell continuation of $s,$ $t,$
and $u$
\begin {eqnarray}
\label {offshell2}
s & = & E^{2}  \nonumber \\
t & = & -({\bf q-q'})^{2}=-q^{2}-q'^{2}+2{\bf q \cdot q'}   \\
u & = & -({\bf q-q'})^{2}=-q^{2}-q'^{2}-2{\bf q \cdot q'}  \nonumber
\end {eqnarray}
applied to eq.~(\ref{rhob2}).
This, therefore, is the ``natural'' or consistent off-shell
continuation of the Mandelstam variables in the BbS approach.
The occurrence of $q^{2}$ and $q'^{2}$ standing separately
violates the requirements of MCC quasi-separability.
In time-ordered perturbation theory, the scattering
particles are on the mass shell, but off the energy shell.
The lagrangian gives one the interactions at
the vertices (the numerators)
and the denominators arise from the normalization of the $\rho$-meson
wave function.  In this case  $q_{1,2}=(\omega _{q},(+,-){\bf q})$ and
$q_{3,4}=(\omega _{q'},(+,-){\bf q'})$, yielding
\begin{equation}
M^{I=0,2}_{\rho ,B} = (2,-1) 2 g_{\rho}^{2}
\left( \frac {2(m_{\pi}^{2}+\omega_{q}\omega_{q'}+q^{2}+q'^{2}
+{\bf q \cdot q'})}{-({\bf q-q'})^{2}-m_{\rho} ^{2}}
+ \frac {2(m_{\pi}^{2}+\omega_{q}\omega_{q'}+q^{2}+q'^{2}
-{\bf q \cdot q'})}{-({\bf q+q'})^{2}-m_{\rho} ^{2}} \right).
\end{equation}
As in the BbS approach, the interaction will not preserve the MCC.

Thus, in order to fulfill the MCC, one cannot simply use interaction
vertices given by the consistent use of a meson-exchange lagrangian in
conjunction with one or another scattering equation.  One can see that
the $\rho \pi \pi$ vertices for $t$- and $u$-channel exchange will
{\em always} have terms in which $q^{2}$ and $q'^{2}$ appear as a sum,
and therefore destroy any possibility for the interaction to be MCC
quasi-separable.  To construct one which is, one must start with
the fully on-shell amplitude expressed in
manifestly crossing-symmetric form , {\em e.g.} eq.~(\ref{rhob2}),
and continue the kinematic variables off shell according to the
prescription of section 2.2 -- or some similar one -- in
which each term in the numerator is either an external energy or mass
squared, or is a product of an initial energy or momentum and a final
one.

\section*{3. Two Models: Numerical Results}

In this section we will examine the results of two models in which the
potential is constructed according to the prescription in section 2.
The results will be compared with those
for the same models when a frequently-used
off-shell continuation which violates MCC quasi-separability
is used.

\subsection*{3.1 The Linear $\sigma$-Model}

We choose the linear $\sigma$-model as a first example.  In this case
we retain the $t$- and $u$-dependence arising from the exchange of
$\sigma$ mesons in the crossed channels. The fit to the $JI$=00 $\pi\pi$
phase shifts is displayed in fig.~1. The parameters and form
factors used for this model are shown in Table I.
The in-medium results, shown in the upper panel of fig.~2,
are quite similar to those of I and confirm the notion that
separability is not the crucial  property of the potential.
(The definition of $M$ differs from that of I by a factor $4\pi^2$.)
The amplitude shows no tendency towards pair condensation,
and there is hardly any invasion of strength into the subthreshold
region even at densities as high as 2$\rho _0$.

In order to demonstrate the consequences of violation of the MCC, we use
the same model, but with the off-shell continuation given by
eq.~(\ref{offshell2}),
the natural one for the BbS reduction of
the Bethe-Salpeter equation, wherein the intermediate pions are on the
energy shell, but off the mass shell. The occurrence in $t$ and $u$ of
$q^{2}$ and $q'^{2}$ standing separately spoil the MCC
quasi-separability of the potential and result in a contribution
from the rescattering integral
which is ${\cal O}(1)$ in the chiral limit.  Results for this model are
shown in the lower panel of fig.~2.
Note that at zero density the amplitude for this
model and for the previous one are very close without having changed the
parameters.  (This good agreement, however, relies on choosing a rather
small bare $\sigma$-mass $m_\sigma$, which allows for a small cutoff
parameter $\Lambda_\sigma$; for higher $\sigma$ masses, which require
higher cutoffs for fitting the $s$-wave phase shifts, the different
off-shell behavior of eqs.~(7) and~(14) induces larger differences.)
  At higher densities, however, one clearly recognizes a much stronger
accumulation of strength below the 2$\pi$ threshold compared to
the model including the MCC (compare upper panel of fig.~2),
although there is no
pair instability in either case at these densities.

\subsection*{3.2 The J\"ulich Model}

While the linear $\sigma$-model is instructive, we wish to examine the
results for a model which achieves a quantitatively better fit to
elastic $\pi \pi$ data: the J\"ulich model~\cite{Loh},
which is based on explicit meson exchange. In order that the
Born amplitude satisfies the soft-pion theorems we here include
contact terms in the lagrangian that arise from a proper gauging
procedure of the non-linear $\sigma$ model~\cite{Wei}. To further
ensure the MCC we employ the on-mass-shell prescription, eq.~(7), for
the off-shell continuation of the pseudopotentials. With some
readjustment of the meson exchange parameters and typical cutoff
parameters $\Lambda_c = 700 - 1000$~MeV for the additionally
introduced contact terms (compare table 2) a good overall fit to
the $\pi\pi$ scattering data can be obtained for c.m. energies well
beyond 1~GeV (see fig.~3). By construction, the $s$-wave scattering
lengths  vanish in the chiral limit (upper panel of fig.~4).
The density dependence of the
amplitude in the $JI$=00-channel is displayed in fig.~5 (upper panel).
In contrast
to the BbS-J\"ulich model employed in I (without contact interactions
and without MCC), in which condensation occurred just above $\rho_0$,
our chirally improved model shows a somewhat weaker tendency towards
instability; as a consequence of the additional repulsion
induced by the contact interactions, the critical density for
condensation is pushed up to approximately 1.4$\rho_0$,
with a moderate
portion of strength absorbed in the condensing peak.  The critical
density can be increased to approximately 2$\rho_0$ with a relatively
small increase in the cutoff mass for the contact terms while retaining
a good fit to the phase shift.
Thus the conjecture that an interaction which produces an amplitude
that satisfies the MCC would eliminate the pair condensation instability
is not correct, even though it appears to work in that direction.

To examine the impact of the MCC in more detail, we replace the
on-mass-shell prescription for the Mandelstam variables, eq.~(7),
by the BbS (on-energy-shell) identification, eq.~(14). A slight
readjustment of some  parameters is necessary to obtain
an overall fit to the $\pi\pi$ phase shifts of similar quality
to that of fig.~3 (see also table 3), although the cutoffs for the
contact terms are kept fixed.  The results are presented in the
lower panel of fig.~5.
The contrast with the previous case is significant.  Condensation
density  is slightly higher -- about 1.5$\rho_0$ -- and can easily
be increased to above 2$\rho_0$ with a small increase of the
contact term cutoffs.
The accumulation of strength in the condensing peaks is very much
smaller than in the previous case and similar to that found for the
J\"ulich model with chiral constraints investigated in I.
  The scattering  lengths as a function of pion mass
(lower panel of fig.~4) show clearly the violation
of the chiral constraints for this model.  Therefore the conjecture
in I that scattering length constraint insures
against pair condensation at low-to-moderate density appears
to be neither sufficient nor necessary.

That is not to say that chiral invariance is unimportant.  The
"improvements" in the J\"ulich model are just those which make the
kernel satisfy the soft-pion theorems.  What is clear now is that
the suppression of the instability is due in large part to the change
in the kernel; there is no necessity to redefine the 2-pion propagator
in order to avoid the instability at near-nuclear density.
One must simply be careful in
constructing the kernel to build in enough repulsion at sub-threshold
energy to suppress the tendency towards $\pi \pi$ bound state formation.
The results are somewhat sensitive to the off-shell continuation of
the kinematic variables, but this sensitivity can largely be compensated
by readjustment of parameters in standard models.

One other difference between the MCC-compliant and BbS cases is worthy
of mention.  Despite the earlier onset of the pairing instability,
the MCC-compliant amplitude actually shows considerably less strength
in the near sub-threshold region, $E\approx1-2m_\pi$, than the BbS case.
This means that at normal nuclear matter
density the range of the in-medium
nucleon-nucleon interaction will be less affected in the MCC case than
in the BbS.  Since the effects are highly non-linear with density, they
are very model-dependent, not only on the $\pi \pi$ interaction, but
also on the approximations underlying the many-body aspects of the
calculation.   Conclusions concerning
the range of the nucleon-nucleon interaction must be made with
extreme care.  Nevertheless, the question of whether the chiral
constraint in some way "protects" the range of the nucleon-nucleon
force up to nuclear matter density is worthy of further study.

\section*{5. Summary}

We have given a general prescription for the construction of $\pi \pi$
potentials for scattering equations which enforces constraints on the
scattering lengths in the limit of zero pion mass, in accordance with
the requirements of chiral symmetry.  When used in a particular
scattering equation, the constrained amplitudes
do not automatically ensure that the $\pi \pi$ interaction in
nuclear matter will be stable against  $\pi \pi$ bound state formation
at low-to-moderate density.  The imposition of the scattering length
constraint  on the kernel does delay the onset of the instability to
higher densities, however; imposition of the constraint on the full
amplitude appears especially effective in suppressing  the $\pi \pi$
interaction strength in the near sub-threshold region.

Due to the complexity of the calculation and the non-linearity of
the in-medium effects, it is difficult to make very general
quantitative statements about the interactions of pions in the nuclear
medium.  Based on our limited investigations it appears that commonly
employed models of the $\pi \pi$ interactions, adjusted to satisfy
the soft-pion theorems, do not necessarily lead to instabilities
when carried over to in-medium calculations.  Small adjustments in
model parameters allow one simultaneously to fit $\pi \pi$ scattering
phases and to avoid the pairing instability at densities below
approximately 2$\rho_0$.  Reformulation of the 2-pion propagator,
which is essentially a definition of a new scattering equation, is
not necessary.  Finally, there are numerical indications that
chiral constraints play a significant part in our understanding of the
nucleon-nucleon interaction in the nuclear medium, but exactly what
that part is remains a subject for further investigation.

\vskip1cm

\centerline {\bf ACKNOWLEDGMENTS}
We are grateful for productive conversations with J. Speth
and K. Nakayama.
One of us (JWD) wishes especially to thank Prof. Speth for
his hospitality
and support during JWD's visits to the Forschungszentrum J\"ulich.
One of us (RR) acknowledges financial support from the German Academic
Exchange Service (DAAD) under program HSPII/AUFE.
This work is supported in part by the National Science Foundation
under Grant No. NSF PHY94-21309.

\pagebreak

\begin{table}
\begin{tabular}{ccccc}
meson &  coupling  &  mass  &  cutoff & form  \\
 X  &  &  $m_X$ [MeV] & $\Lambda_X$ [MeV] & factor \\
\hline
s-channel $\sigma^{(0)}$
& \quad $(m_\sigma^2-m_\pi^2)/f_\pi^2 \  \vec\pi^2 \sigma$  \quad &
 800 & 1250 & $(2\Lambda^2)^2/(2\Lambda^2+4q^2)^2$
\\
t-channel $\sigma \quad $ & " &  " & " &
$(2\Lambda^2)^2/(2\Lambda^2+(\vec q - \vec{q'})^2)^2$  \\
\end{tabular}
\caption{Coupling, parameters and form factor types
used in the linear $\sigma$-model
for both on-energy-shell and on-mass-shell prescription}
\end{table}
\begin{table}
\begin{tabular}{ccccc}
$\pi\pi$ channel  \\
\hline \hline
meson &  coupling  & coupl. const.&
 mass  &  cutoff  \\
 X  &  & $(g_{\pi\pi X})^2/4\pi$ & $m_X$ [MeV] & $\Lambda_X$ [MeV] \\
\hline
$\epsilon^{(0)}$(1400)  &
$(g_{\pi\pi\epsilon}/m_\pi)  \ (\partial_\mu \vec{\pi} \cdot
\partial^\mu \vec{\pi}) \phi_\epsilon$  \quad &
  0.008   &   1585  &   1750
\\
$\rho^{(0)}$(770) & $g_{\pi\pi\rho}(\vec \pi \times
\partial_\mu\vec \pi) \cdot \vec \rho^\mu$ &
1.6 & 1045 &  3000
\\
 $\rho$(770) & " &
3.0 & 770 & 1410
\\
$f_2^{(0)}$(1270) & $(g_{\pi\pi f_2}/m_\pi)  \ (\partial_\mu \vec \pi
\cdot \partial_\nu  \vec \pi) \phi_{f_2}^{\mu\nu} $
& 0.015 &  1573 & 2285
\\
\hline
contact & $(m_\pi^2/8f_\pi^2) \  (\vec\pi^2)^2$  & $f_\pi$=93 MeV &
 $m_\pi$=139.57  & 700
\\
inter- & $(1/4f_\pi^2) \ (\vec\pi)^2 ( \partial_\mu\vec\pi  \cdot
\partial^\mu\vec\pi)$ & " & --  & "
\\
 actions & $(g_{\pi\pi\rho}^2/2m_\rho^2) \
 (\vec\pi\times\partial^\mu\vec\pi)^2$
 & 1.6 & $m_\rho$=770 & "
\\
& & & &
\\
\hline\hline
$\pi\pi \rightarrow K\bar K$ channel
\\
\hline \hline
meson &  coupling  & coupl. const.&
 mass  &  cutoff  \\
 X &  & $(g_{K\pi X})^2/4\pi$ & $m_X$ [MeV] & $\Lambda_X$ [MeV] \\
\hline
$K^*$(895) & $g_{\pi K K^*}\partial^\mu\vec\pi \cdot (K \vec \tau
 K_\mu^*)$ &
0.75 & 895 & 1410
\\
 & & & &
\\
\hline\hline
$K\bar K$ channel  \\
\hline \hline
meson &  coupling  & coupl. const. &
 mass  &  cutoff  \\
  X &  & $(g_{K\bar K X})^2/4\pi$ & $m_X$ [MeV] & $\Lambda_X$ [MeV] \\
\hline
$\epsilon^{(0)}$(1400)  &
$(g_{K\bar Kf_0}/m_K) \partial_\mu \bar K
\partial^\mu K \phi_{f_0}$  \quad &
 0.002 & 1585 & 1750
\\
$\rho^{(0)}$(770) & $g_{K\bar K\rho}( K \vec\tau \partial_\mu{\bar K})
\cdot \vec \rho^\mu$ &
0.4 & 1045  & 3000
\\
 $\rho$(770) & "  &
0.75 & 770 &  2135
\\
$\omega$(782) & $g_{K\bar K\omega}( K  \partial_\mu{\bar K})
 \omega^\mu$ &
--0.75 & 782.6  &  2135
\\
$\phi$(1020) & $g_{K\bar K\phi}( K  \partial_\mu{\bar K})
 \phi^\mu$ &
--1.5 & 1020 &  2135
\\
$f_2^{(0)}$(1270) & $(g_{K\bar Kf_2}/m_K) \partial_\mu K \partial_\nu
{\bar K}\phi_{f_2}^{\mu\nu} $ &
0.004 & 1573 & 2285
\\
\end{tabular}
\caption{Couplings and parameters of the chirally improved
$\pi\pi$/$K\bar K$ J\"ulich model employing the on-mass-shell
prescription, eq.~(7),
for the off-shell continuation of the pseudopotentials;
s-channel pole graphs are labelled with superscript '(0)';
form factors used are of dipole type: $F^{(s)}(q)$=$(2\Lambda^2+m^2)^2/
(2\Lambda^2+4\omega_q^2)^2$ for s-channel pole graphs,
$F^{(t)}(q,q')$=$(2\Lambda^2-m^2)^2/(2\Lambda^2+(\vec q -\vec{q'})^2)^2$
for t-channel exchange graphs and
$F^{(c)}(q)$=$(2\Lambda^2-4m_\pi^2)^2/(2\Lambda^2+4q^2)^2$ for contact
interactions.}
\end{table}
\begin{table}
\begin{tabular}{ccccc}
$\pi\pi$ channel  \\
\hline \hline
meson &  coupling  & coupl. const.&
 mass  &  cutoff  \\
 X  &  & $(g_{\pi\pi X})^2/4\pi$ & $m_X$ [MeV] & $\Lambda_X$ [MeV] \\
\hline
$\epsilon^{(0)}$(1400)  &
$(g_{\pi\pi\epsilon}/m_\pi)  \ (\partial_\mu \vec{\pi} \cdot
\partial^\mu \vec{\pi}) \phi_\epsilon$  \quad &
  0.016   &   1545  &   1900
\\
$\rho^{(0)}$(770) & $g_{\pi\pi\rho}(\vec \pi \times \partial_\mu\vec
\pi)  \cdot \vec \rho^\mu$ &
1.6 & 1108 &  3500
\\
 $\rho$(770) & " &
3.0 & 770 & 1450
\\
$f_2^{(0)}$(1270) & $(g_{\pi\pi f_2}/m_\pi)  \ (\partial_\mu \vec \pi
\cdot \partial_\nu  \vec \pi) \phi_{f_2}^{\mu\nu} $
& 0.015 &  1710 & 2770
\\
\hline
contact & $(m_\pi^2/8f_\pi^2) \  (\vec\pi^2)^2$  & $f_\pi$=93 MeV &
 $m_\pi$=139.57  & 700
\\
inter- & $(1/4f_\pi^2) \ (\vec\pi)^2 ( \partial_\mu\vec\pi  \cdot
\partial^\mu\vec\pi)$ & " & --  & "
\\
 actions & $(g_{\pi\pi\rho}^2/2m_\rho^2) \
 (\vec\pi\times\partial^\mu\vec\pi)^2$
 & 1.6 & $m_\rho$=770 & "
\\
& & & &
\\
\hline\hline
$\pi\pi \rightarrow K\bar K$ channel
\\
\hline \hline
meson &  coupling  & coupl. const.&
 mass  &  cutoff  \\
 X &  & $(g_{K\pi X})^2/4\pi$ & $m_X$ [MeV] & $\Lambda_X$ [MeV] \\
\hline
$K^*$(895) & $g_{\pi K K^*}\partial^\mu\vec\pi \cdot (K \vec \tau
 K_\mu^*)$ &
0.75 & 895 & 1575
\\
 & & & &
\\
\hline\hline
$K\bar K$ channel  \\
\hline \hline
meson &  coupling  & coupl. const. &
 mass  &  cutoff  \\
  X &  & $(g_{K\bar K X})^2/4\pi$ & $m_X$ [MeV] & $\Lambda_X$ [MeV] \\
\hline
$\epsilon^{(0)}$(1400)  &
$(g_{K\bar Kf_0}/m_K) \partial_\mu \bar K
\partial^\mu K \phi_{f_0}$  \quad &
 0.004 & 1545 & 1900
\\
$\rho^{(0)}$(770) & $g_{K\bar K\rho}( K \vec\tau \partial_\mu{\bar K})
\cdot \vec \rho^\mu$ &
0.4 & 1108  & 3500
\\
 $\rho$(770) & "  &
0.75 & 770 &  2275
\\
$\omega$(782) & $g_{K\bar K\omega}( K  \partial_\mu{\bar K})
 \omega^\mu$ &
--0.75 & 782.6  &  2275
\\
$\phi$(1020) & $g_{K\bar K\phi}( K  \partial_\mu{\bar K})
 \phi^\mu$ &
--1.5 & 1020 &  2275
\\
$f_2^{(0)}$(1270) & $(g_{K\bar Kf_2}/m_K) \partial_\mu K \partial_\nu
{\bar K}\phi_{f_2}^{\mu\nu} $ &
0.004 & 1710 & 2770
\\
\end{tabular}
\caption{Same as table~2, but employing the BbS (on-energy-shell)
prescription, eq.~(14).}
\end{table}
\vfill\eject

\newpage
\begin{center}
{\large \sl \bf Figure Captions}
\end{center}
\vspace{0.5cm}

\begin{itemize}
\item[{\bf Figure 1}:] Fit to the scalar-isoscalar $\pi\pi$ phase
shifts with the linear $\sigma$-model. The dashed line is the
result when employing the on-mass-shell prescription,
eq.~(\ref{offshell1}), whereas the full line
corresponds to the on-energy-shell (BbS) prescription,
eq.~(\ref{offshell2}), using the same set of parameters (see table I).

\item[{\bf Figure 2:}] Scalar-isoscalar invariant $\pi\pi$
$M$ amplitude with the linear $\sigma$ model in nuclear matter;
long-dashed lines: $\rho/\rho_0$=0.5, short-dashed lines:
$\rho/\rho_0$=1.0, dotted lines: $\rho/\rho_0$=2.0; the full lines
correspond to the amplitude in free space. \\
The upper panel shows the results using the on-mass-shell
prescription, whereas the lower panel corresponds to the BbS
prescription.

\item[{\bf Figure 3:}] Fit to the $\pi\pi$ phase shifts with
the J\"ulich model supplemented by contact interactions and
using the on-mass-shell prescription; parameters used are
listed in table II.   \\
A very similar fit is obtained using the BbS prescription with
the slightly modified parameter set of table III (contact
term parameters unchanged).

\item[{\bf Figure 4:}] $S$-wave $\pi\pi$ scattering lengths in the
chiral limit for various versions of the J\"ulich model; \\
upper panel: on-mass-shell prescription including contact interactions
(corresponding to the model of fig.~3 / table II);  \\
lower panel: BbS prescription, dashed lines: including contact
interactions (corresponding to the model of table III),
dotted lines: without contact interactions.

\item[{\bf Figure 5:}] Scalar-isoscalar invariant $\pi\pi$ amplitude
in nuclear matter for the J\"ulich model supplemented with contact
interactions;  \\
upper panel: employing the on-mass-shell prescription (corresponding
to table II), long-dashed line: $\rho/\rho_0$=0.5, short-dashed line:
$\rho/\rho_0$=1.0, dotted line: $\rho/\rho_0$=1.3;  \\
lower panel: employing the BbS prescription, line identification as
in upper panel except that the dotted
line is for $\rho/\rho_0$=1.45. \\
The full lines correspond to the amplitudes in free space.

\end{itemize}

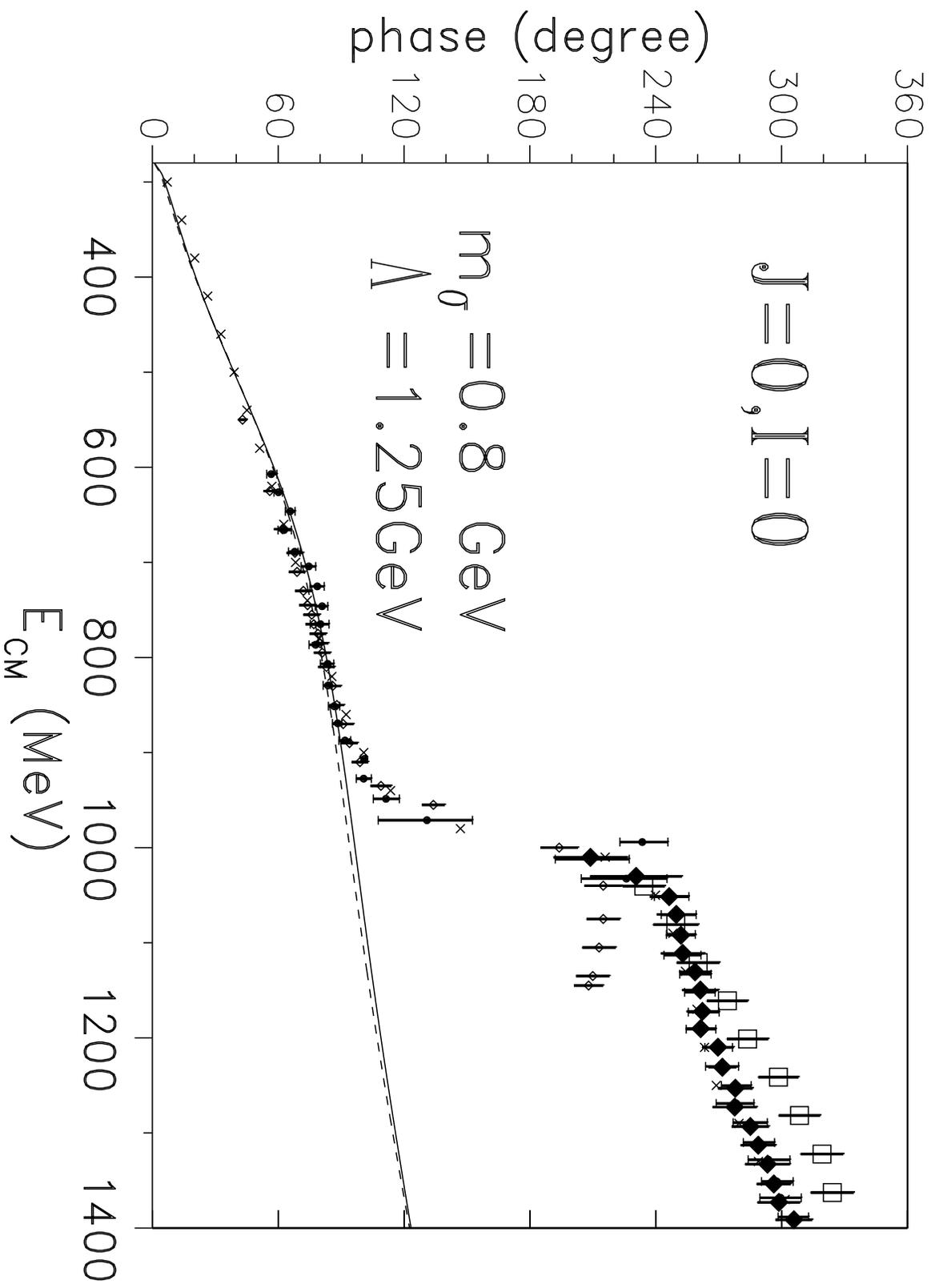
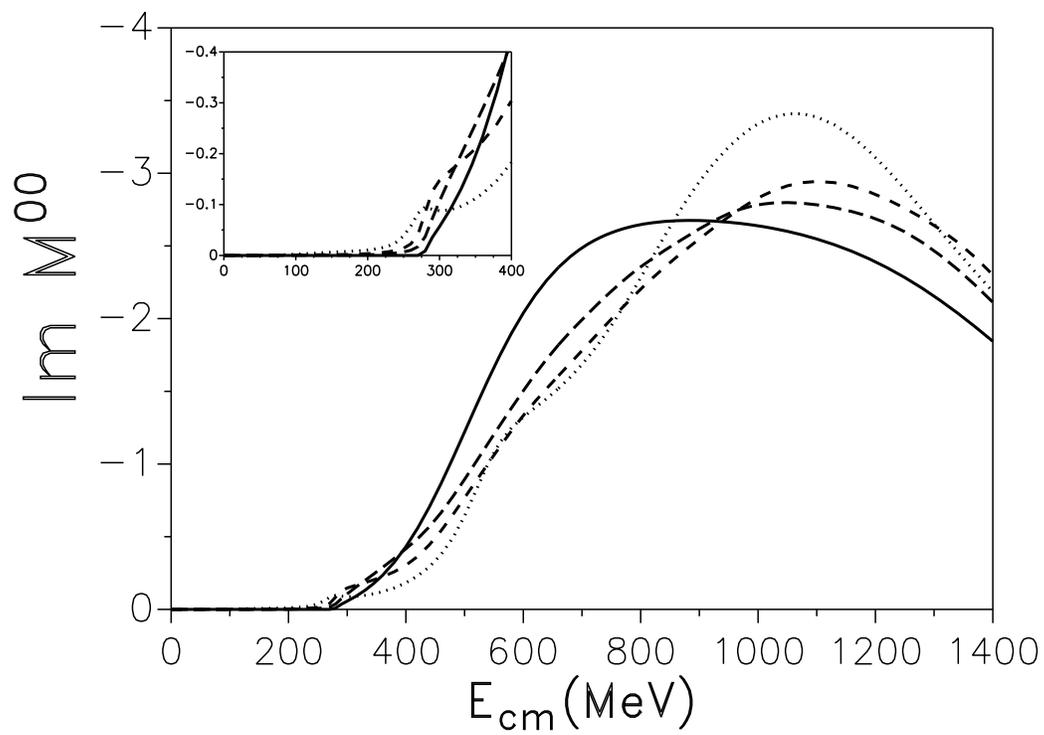
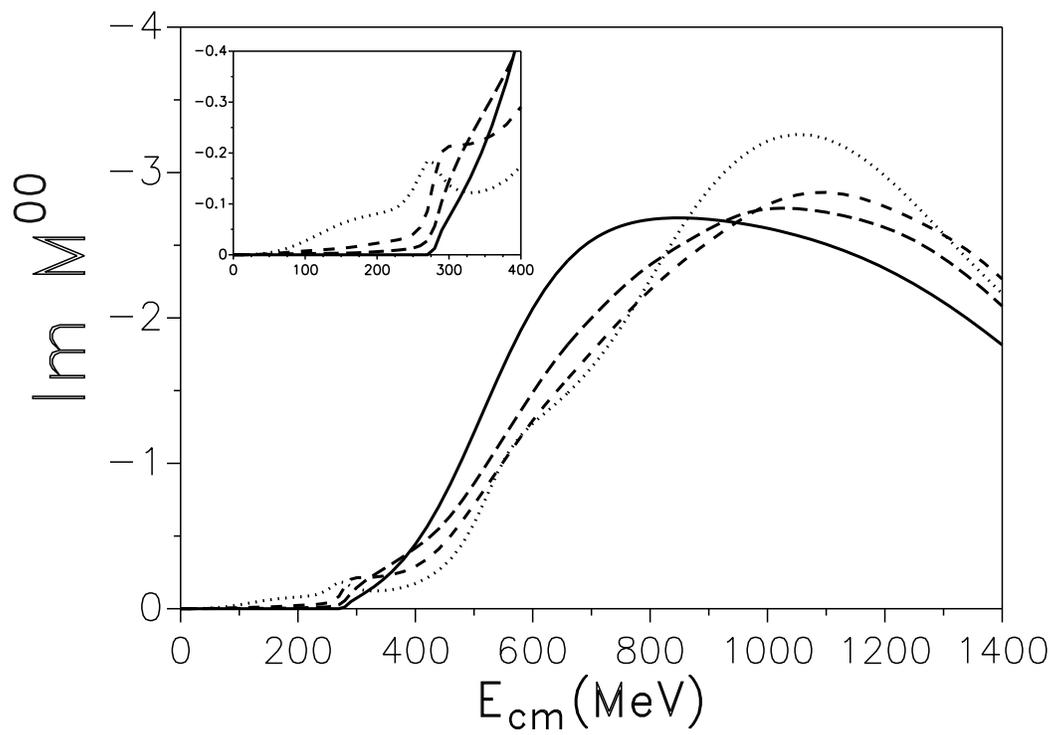
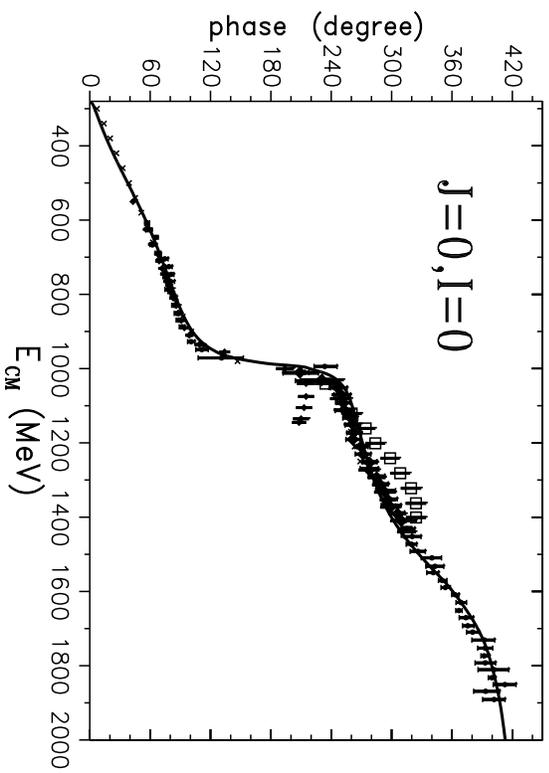
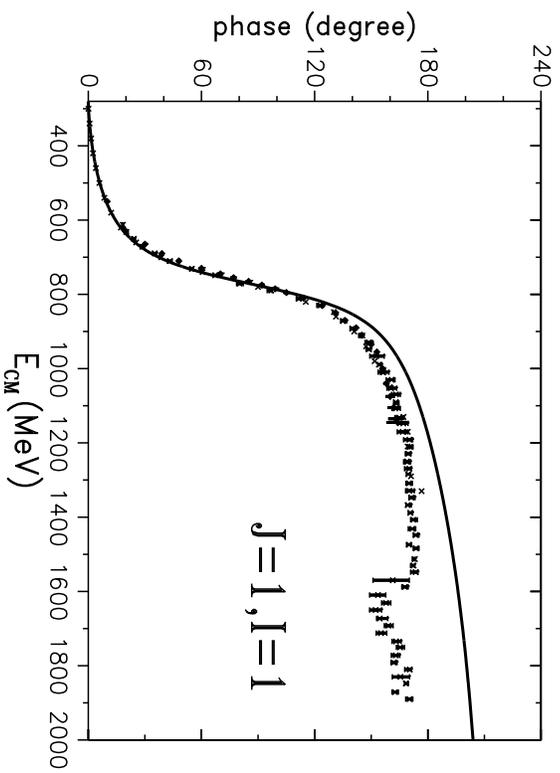
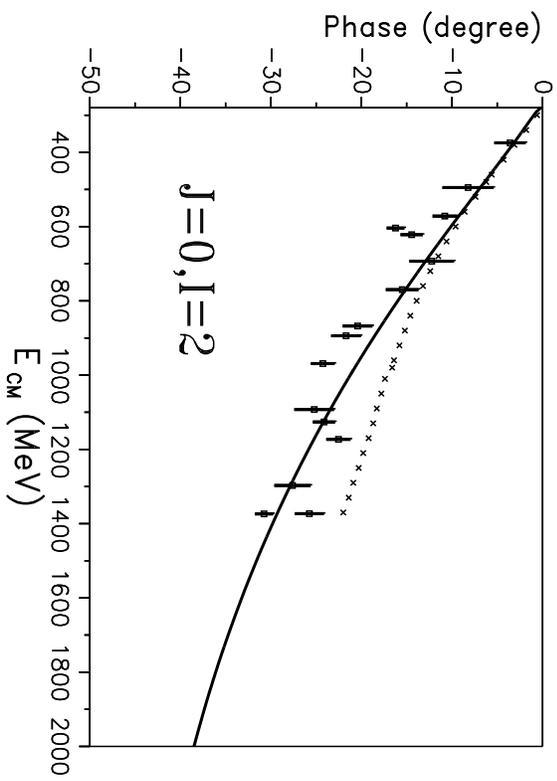
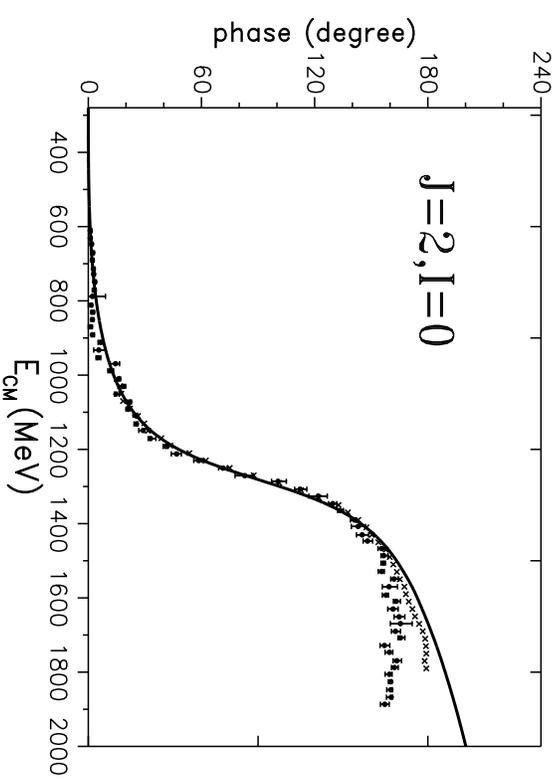
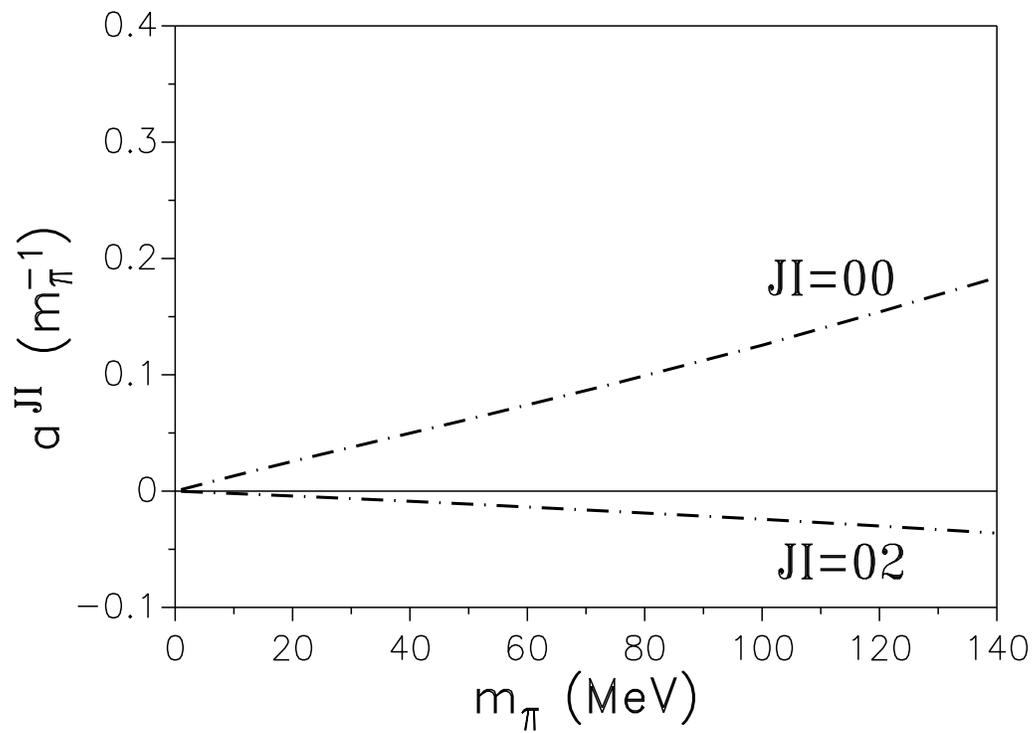
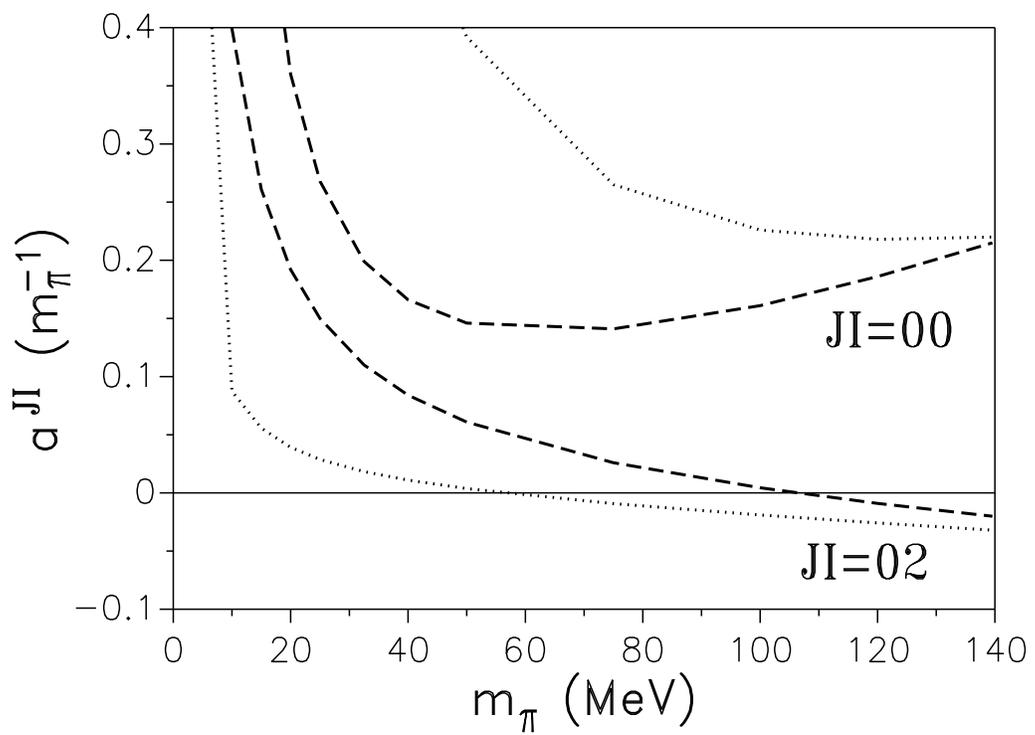
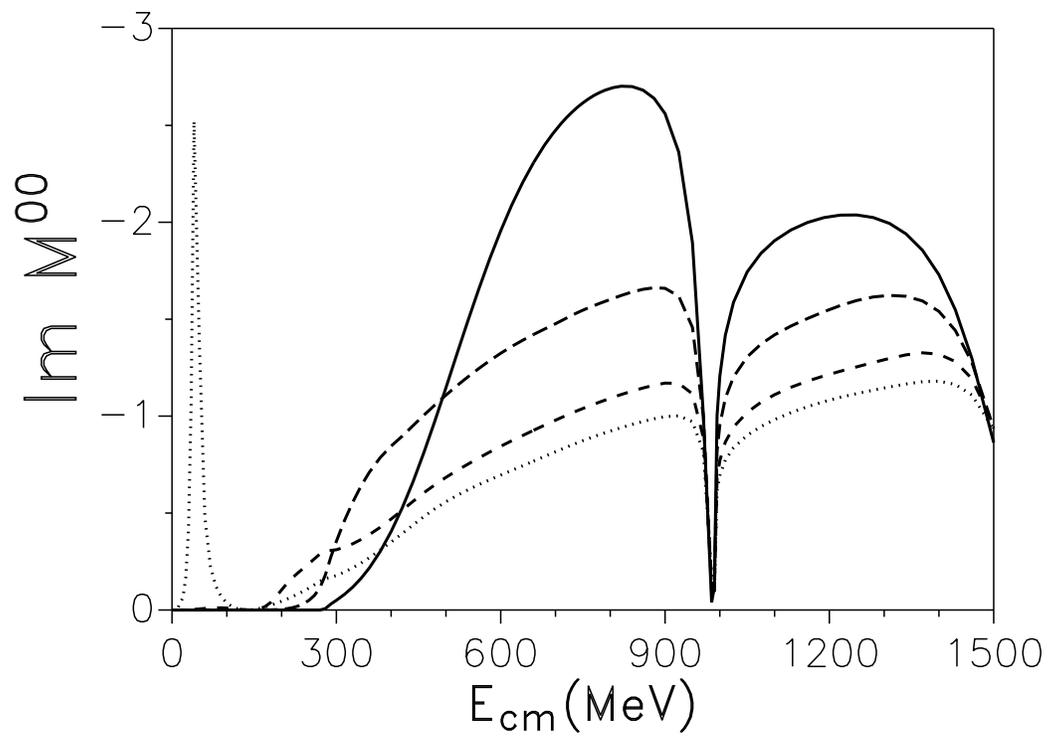
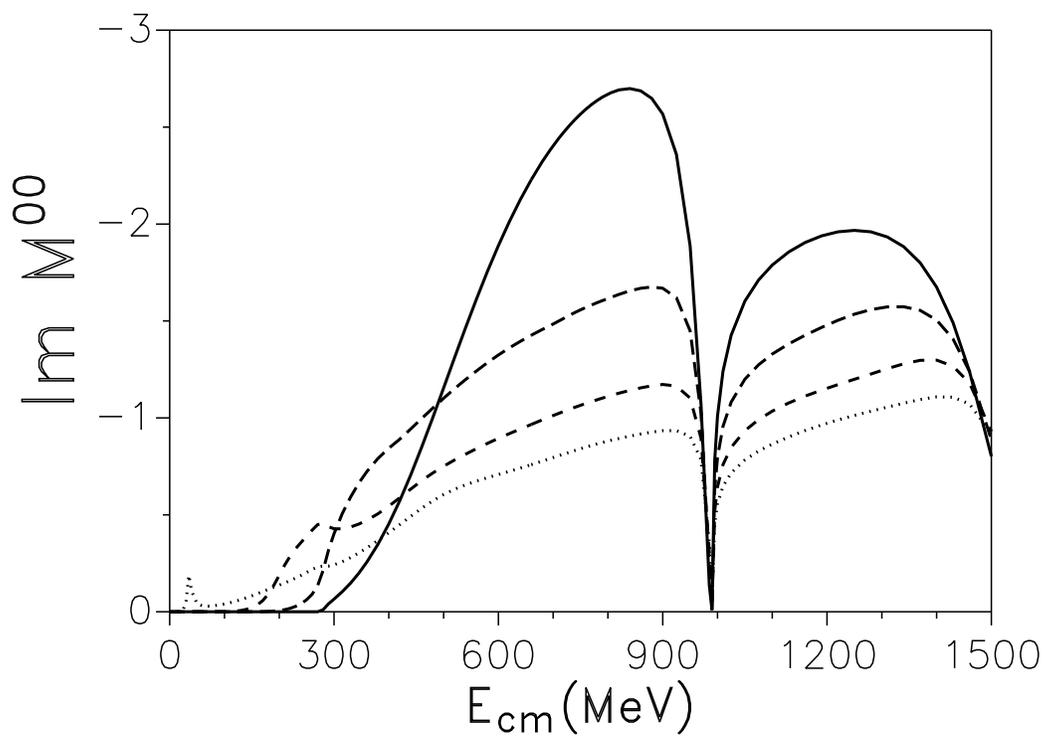
\end{document}